\documentclass[superscriptaddress]{revtex4-2}

\usepackage{amssymb}
\usepackage{amsmath}
\usepackage{revsymb}
\usepackage{graphicx}
\usepackage{xcolor}
\usepackage{hyperref}

\definecolor{urlcolor}{rgb}{0,.2,.5}
\definecolor{linkcolor}{rgb}{0,.15,.7}
\definecolor{citecolor}{rgb}{.12,.54,.11}

\hypersetup{breaklinks=true,colorlinks=true,urlcolor=urlcolor,linkcolor=linkcolor,citecolor=citecolor}

\begin{document}

\title{Monte Carlo Simulations of Suprathermal Enhancement in Advanced Nuclear Fusion Fuels}

\author{Marcus Borscz}
\email{marcus.borscz@hb11.energy}
\affiliation{HB11 Energy Holdings Pty Ltd, Sydney, NSW, Australia}
\affiliation{School of Mechanical and Manufacturing Engineering, The University of New South Wales, Sydney, NSW, Australia}

\author{Thomas A.\ Mehlhorn}
\affiliation{HB11 Energy Holdings Pty Ltd, Sydney, NSW, Australia}

\author{Patrick A.\ Burr}
\affiliation{School of Mechanical and Manufacturing Engineering, The University of New South Wales, Sydney, NSW, Australia}
             
\author{Igor Morozov}
\affiliation{HB11 Energy Holdings Pty Ltd, Sydney, NSW, Australia}
\affiliation{School of Mathematical and Physical Sciences, Macquarie University, Sydney, NSW, Australia}

\author{Sergey Pikuz}
\affiliation{HB11 Energy Holdings Pty Ltd, Sydney, NSW, Australia}


\date{\today}

\begin{abstract}
Suprathermal fusion reactions, initiated by energetic particles slowing down and scattering in dense plasmas, can modify the burn dynamics at inertial confinement fusion (ICF) regimes. A 0D time-dependent Monte-Carlo code has been developed to assess the suprathermal energy gain from fast fusions in DT, deuterium, $^{11}$BH$_3$ and $^{11}$BHDT fuels. It incorporates modified Li–Petrasso stopping powers, thermal broadening of cross-sections, anisotropic nuclear elastic and neutron elastic scattering, and a physical model for the p$^{11}$B alpha-particle spectra. Results show that earlier predictions of suprathermal criticality in pure deuterium are overestimated by more than an order of magnitude; no realistic density–-temperature regime supports a self-sustaining chain reaction. Fast protons in $^{11}$BH$_3$ have an optimum energy of 4 MeV for maximising suprathermal enhancement. In this case the additional energy from fast fusions is unlikely to exceed 40\% of the initial proton beam energy. The possibility of an alpha-particle-driven ‘avalanche’ mechanism is ruled out since the ionic stopping is dominated by collisions involving small energy transfer. Suprathermal multiplication processes are dominated by neutron-driven ion up-scattering and likely play a limited role in purely aneutronic fuels.
\end{abstract}

\maketitle

\section{Introduction}
The considerable achievement of scientific breakeven at the National Ignition Facility (NIF) \cite{AbuShawareb:PRL24} marks an exciting new paradigm for fusion research by providing a robust experimental platform to study the kinetic interactions between thermal fuel ions and fast fusion products. One of the surprises from the first burning plasma shots at the NIF is the observation of anomalous neutron spectra that are characteristic of stronger non-Maxwellian ion populations than otherwise predicted by elastic up-scattering of ions by fusion $\alpha$-particles \cite{Hartouni:NP23}.

Simulations with the hybrid particle-in-cell (PIC) code LAPINS \cite{Xue:SB25} were shown to be congruent with the NIF experimental results by accounting for Coulomb large-angle scattering (CLS) using a physically-motivated cutoff for the impact parameter \cite{Turrell:PRL14}. The small impact parameters required for these scattering events are typical of head-on collisions with high relative but low centre-of-momentum (CoM) velocities; their inclusion amplifies the isotropic neutron velocity whilst mitigating spectral broadening. Moreover, the peak density of $\alpha$-particles at the hotspot boundary almost doubles with the inclusion of CLS, which facilitates a more rapid hydrodynamic expansion and more efficient burn propagation at cooler hotspot temperatures for a given yield.

Recently, the effectiveness of CLS to appreciably enhance the yield and modify the neutron spectra has been called into question following development of the PICNIC code by Lawrence Livermore National Laboratory (LNLL) \cite{Wetering:PRE25l}. They employ a rather elegant `generalised Coulomb method' \cite{Angus:JCR25} that computes the large-angle cutoff from the normalised transport mean free path. For ICF-relevant moderately coupled plasmas where the Coulomb logarithm $2\lesssim\ln\Lambda\lesssim 10$, this method reproduces the third-order $\sim 1/\ln\Lambda$ corrections to the Fokker-Planck collision operator in the limit of small time steps \cite{Li93,Zylstra:PP19}. With CLS the D/T ion distributions display a large enhancement in the high-energy tail ($>$100 keV), although these knock-on ions represent a small fraction ($\sim 5\%$) of the total population. The production of suprathermal ions at multi-MeV energies is in any case dominated by nuclear elastic scattering (NES). The neutron spectrum on PICNIC remains broadly consistent with hydrodynamic simulations even with inclusion of CLS, $\alpha$-D/$\alpha$-T NES and anisotropic fusion. At the time of writing the discrepancy between LAPINS and PICNIC is yet to be resolved and an agreed-upon mechanism for the anomalous spectral shifts at the NIF remains elusive. 

It is anticipated that another kinetic regime exists at even higher densities, that of a suprathermal chain reaction driven by elastic recoils \cite{Gryzinski58}. Peres and Shvarts \cite{Peres75} show that cold infinite DT plasma can be made critical at densities above $3.6\times 10^{4}$ g cm$^{-3}$. This can be reduced to $3.1\times 10^{4}$ g cm$^{-3}$ by seeding with $1\%$ $^6$Li, since the (n,$\alpha$) reaction yields 2.06 MeV $\alpha$-particles which happen to fall on the $\alpha$-D NES resonance. For a finite temperature plasma, Afek \textit{et al.} \cite{Afek:JPD78} have reported lower densities, for instance $4200$ g cm$^{-3}$ at $T_i=15$ keV (for all following discussion, we assume isothermal electrons and ions unless stated otherwise). Under less extreme conditions more relevant to near-term driver capabilities, the effect of a subcritical suprathermal chain reaction can still provide appreciable enhancement if DT neutrons effectively couple to the ions. For $\rho=250$ g cm$^{-3}$ and $T_i=30$ keV there is a 50\% probability of a DT fusion per 14.1 MeV neutron even in the conservative case of a $3T_i$ energy cutoff \cite{Kumar87}. The usual practice of ignoring neutrons in burn models \cite{Atzeni:book04} may need to be revisited to account for this burn enhancement at high density.

\begin{figure}[b]
 \centering
        \includegraphics[width=0.5\linewidth]{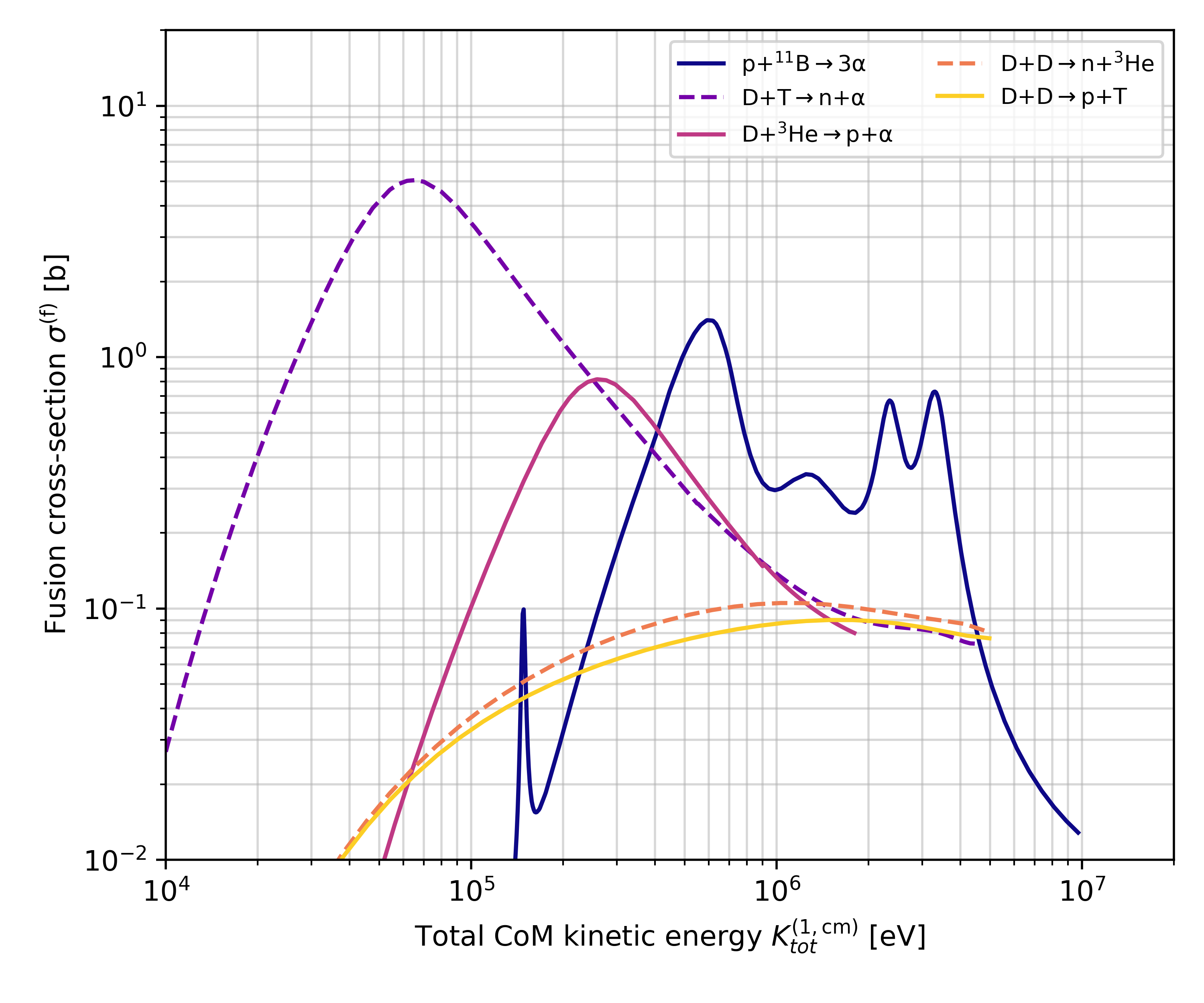}
 \caption{Fusion cross-sections of candidate fuels. Reactions involving deuterium are
taken from Bosch and Hale \cite{Bosch92}, the p$^{11}$B cross-section is taken from Tentori and Belloni \cite{Tentori23}.}
\label{fig1}
\end{figure}

Interest in suprathermal chain reactions has been renewed in context of so-called `advanced' fusion fuels \cite{Nevins:JFE98}, which most notably include pure deuterium, D$^3$He and p$^{11}$B. Compared to DT, these fuels exhibit reduced neutron production, more available fuel supply and potentially simpler reactor engineering, at the expense of less favourable cross-sections and higher radiative losses. Figure \ref{fig1} compares the cross-sections of these candidate fuels with DT.
A notable recent development is the claim by Robinson \cite{Robinson24a, Robinson24b} that a chain reaction in pure deuterium can proceed at $\rho=3300$ g cm$^{-3}$ and $T_i=25$ keV when initiated by a 6 MeV deuteron beam. In fact the critical density for degenerate deuterium was found to be $1.2\times 10^{4}$ g cm$^{-3}$, which is three times less than what is required for DT. Here the beam-target gains were obtained using a Monte Carlo model that samples fusion, neutron elastic scattering and p-D elastic scattering cross-sections as fast particles slow down in a thermal background. Crucially, the chain reaction mechanism is only possible because of the up-scattering of deuterons by high-energy neutrons from DT fusions, which follow from the D(d,p)T reaction branch.

Multiplication processes are likely to be essential for any fusion scheme using low-neutronicity p$^{11}$B \cite{Belloni:LPB22}. Perhaps the most rigorous exploration of these effects was by Weaver \textit{et al.} \cite{Weaver:Preprint73}, who in 1973 suggested a 5--15\% suprathermal enhancement using relativistic Fokker-Planck simulations \cite{Zimmerman76} with inclusion of CLS and NES. However, the old cross-section data was used and further open-access details remain limited. Putvinski \textit{et al.} \cite{Putvinski:NF19} demonstrated that the parameter space for p$^{11}$B self-heating is marginal after incorporating the re-evaluated cross-section \cite{Sikora:JFE16} and kinetic modification of the proton distribution from $\alpha$-particle slowing down. The latter contributes an additional 10\% to the reactivity at magnetic fusion conditions. By considering both studies, subsequent 0D \cite{Ghorbanpour:FP24} and 1D \cite{Borscz:Preprint25_2} burn models have assumed a maximum enhancement of 20\% due to deviations from Maxwellian ion distributions. A more speculative `avalanche' process \cite{EliezerHora:PP16} driven by $\alpha$-p elastic scattering was proposed to explain unexpectedly high yields in laser-plasmas produced by boron-hydrogen-silicon targets \cite{Picciotto:PRX14}. This has since been ruled out by analytic studies \cite{Belloni:PPCF21, Hirose:PPCF25} which show that the $\alpha$-multiplication factor does not exceed a few percent even at extreme conditions ($\rho=5\times10^4$ g cm$^{-3}$, $T_i=200$ keV). Indeed commercial p$^{11}$B fusion, if viable, will likely require a thermonuclear approach to achieve ignition, and proton beam fast ignition is particularly advantageous because additional heating is provided by in-flight reactions \cite{Mehlhorn:LPB22}. Moreau \cite{Moreau:NF77} estimates that it is difficult to achieve an energy multiplication factor greater than 30\%, maximised at 4 MeV proton injection energy. These estimates should be revisited in light of the updated cross-section and more realistic stopping models that include large-angle scattering and plasma dielectric effects.

In this paper we substantially extend Robinson's Monte Carlo model \cite{Robinson24a} by incorporating an updated stopping power, thermal motion effects in the collision kinematics, a physical model for the p$^{11}$B $\alpha$-spectrum and anisotropic elastic scattering. Validation against an analytic model for pure deuterium confirms that no practically achievable plasma attains criticality, contrary to earlier claims. We extend Moreau's \cite{Moreau:NF77} results for p$^{11}$B beam-target fusion to ICF conditions and show only a slight increase in the maximum gain. $\alpha$-particles are highly ineffective for driving suprathermal reactions, although fast neutrons can produce an enhancement that is comparable to that of protons. We appreciate that the most important feature of pure p$^{11}$B, its low neutronicity, may in fact work against the exploitation of suprathermal processes, so we briefly examine the suprathermal energy multiplication in mixed p$^{11}$B and DT fuel. While we do not comment on the ignition requirements of these mixed fuels, we show that 14.1 MeV neutrons scatter strongly off protons and can produce an enhancement effect on the order of 30\%. We conclude by evaluating the suprathermal enhancement for thermal DT, pure deuterium, $^{11}$BH$_3$ and $^{11}$BHDT in $\rho$--$T_i$ space.

\section{Model}
A numerical Monte Carlo model has been developed to track fast neutrons and ions as they undergo collisions with an infinite and homogeneous thermal plasma. In the current implementation the only ions considered are protons, deuterons, tritons, helions, $\alpha$-particles and $^{11}$B, which covers the main candidate reactions for fusion energy.

In a simulation, a primary particle initialised with kinetic energy $E_p$ initiates a collision chain of suprathermal particles $s$, which may include the primary as well any daughters produced via scattering or reactions. Each suprathermal particle is defined by its parent collision chain, its mass $m_s$, charge number $Z_s$ and kinetic energy $K_s$, and exists at time $t$. During each simulation step, suprathermal particles first lose energy due to the electronic and ionic stopping power (if charged) before undergoing a discrete collision with a sampled plasma species. Of particular interest is the suprathermal energy gain, defined for the primary particle as $G=\mathcal{K}/E_p - 1$, where $\mathcal{K}$ is the total energy deposited to the plasma by the resulting collision chain  (i.e. the kerma). This definition is equivalent to the energy multiplication factor from Moreau \cite{Moreau:NF77} and the beam-target gain from Robinson \cite{Robinson24a}.

\subsection{Continuous slowing down}
We denote a particle’s kinetic energy immediately before slowing down, after slowing down but before a collision, and after a collision as $K_s^{(0,\text{lab})}$, $K_s^{(1,\text{lab})}$ and $K_s^{(2,\text{lab})}$, respectively. In contrast to Robinson's fixed timestep scheme \cite{Robinson24a, Robinson24b}, we adopt a more computationally efficient asynchronous update method in which each particle advances from its current time $t$ depending on its collision rate. The probability of a collision follows an exponential distribution so we can capture the stochastic nature of interactions by sampling the expected number of collisions for each particle, $\Delta N=-\ln(1-u)$, where uniform variable $u\sim U(0,1)$. Neutrons only change their energies at collision events, so $K_s^{(1,\text{lab})}=K_s^{(0,\text{lab})}$ and $\Delta t=\Delta N/\sum_b\bar{\nu}_{sb}$, where $\bar{\nu}_{sb}$ is the Maxwellian-averaged collision frequency with the background ion species $b$, derived in appendix \ref{appA}.

For ions, continuous slowing down requires evaluating the following integrals,
\begin{gather}
    \Delta N=\int_{K_s^{(0,\text{lab})}}^{K_s^{(1,\text{lab})}} \frac{\sum_b \bar{\nu}_{sb}}{v_s}\left(\frac{dK_s}{dx}\right)^{-1}  dK_s, \label{eqn:deltaN} \\
    \Delta t = \int_{K_s^{(0,\text{lab})}}^{K_s^{(1,\text{lab})}} \frac{1}{v_s}\left(\frac{dK_s}{dx}\right)^{-1} dK_s. \label{eqn3}
\end{gather}
Here $v_s=\sqrt{2K_s/m_s}$ is the suprathermal particle speed and $dK_s/dx$ is the stopping power. To avoid repeatedly evaluating these integrals during the Monte Carlo simulation, we precompute them from $10^8$ eV to $3T_i/2\leq K_s\leq 10^8$ and store as tabulated functions $\tilde{N}(K_s)$ and $\tilde{t}(K_s)$. The grid is defined such that the relative error is $<1\%$ at the midpoints. $K_s^{(1,\text{lab})}$ and $\Delta t$ are then computed as follows
\begin{gather}
    K_s^{(1,\text{lab})}=\tilde{N}^{-1}\left(\tilde{N}(K_s^{(0,\text{lab})}) + \Delta N\right), \\ \Delta t=\tilde{t}(K_s^{(1,\text{lab})}) - \tilde{t}(K_s^{(0,\text{lab})}).
\end{gather}

For the stopping power we use the modified parameterisation of the Li-Petrasso model (mLP) from Zylstra \textit{et al.} \cite{Zylstra:PP19},
\begin{gather}
    \frac{dK_s}{dx}=-4\pi \left(\frac{\kappa Z_s}{v_s}\right)^2 \sum_j  \frac{n_j Z_j^2}{m_j} \left(G\left(\frac{v_s}{v_{j,th}}\right)\ln\Lambda_{sj} + H\left(\frac{v_{j,th}}{\sqrt{2}v_s}\right)\right), \label{eqn:mLP}
\end{gather}
where $j$ is taken over both background ion and electron species. Here $\kappa=e^2/4\pi\varepsilon_0$ is the Coulomb coupling constant, $n_j$ is the number density of the target species and $v_{j,th}=\sqrt{2T_j/m_j}$ is its thermal speed. The binary collision term is given by 
\begin{equation}
    G(u)=\left(1+\frac{m_j}{m_s\ln\Lambda_{sj}}\right) \text{erf}(u) - \frac{2}{\sqrt{\pi}}\left(1+\frac{m_j}{m_s}\right)ue^{-u}, \label{eqn:Gterm}
\end{equation}
and includes the third-order $1/\ln\Lambda$ contribution representing energy loss from CLS. The collective term $H(u)=u K_0(u) K_1(u)$, where $K_0$ and $K_1$ are irregular modifed cylindrical Bessel functions, accounts for the dielectric response of the plasma and dominates at low temperatures. We note that most previous studies of suprathermal enhancement \cite{Moreau:NF77, Robinson24a, Belloni:PPCF21, Kumar87} neglect the additional stopping power from CLS and collective effects. The definition for the Coulomb logarithm is,
\begin{equation}
    \ln\Lambda_{sj} = \frac{1}{2}\ln\left(1+\frac{\lambda_D^2}{b_\perp^2+\lambdabar}\right),
\end{equation}
for Debye length $\lambda_D$, 90$^\circ$ impact parameter $b_\perp$ and de Broglie wavelength $\lambdabar$. We point readers to the original paper \cite{Zylstra:PP19} for the calculation of these terms as well as the required electron degeneracy correction. We note that their equations 6 and 7 have been erroneously multiplied by $\sqrt{\pi}/2$ and has been corrected in this work. Our implementation has also been validated for the stopping of deuterons in deuterium, see case (2) in figure 3 of \cite{Ghosh_2007}.

Once the post-slowing down energy is computed, the kerma associated with that particle's parent collision chain is incremented by $K_s^{(0,\text{lab})}-K_s^{(1,\text{lab})}$. If the particle is deemed thermalised then the kerma is instead incremented by $K_s^{(0,\text{lab})}$ and the particle is removed from the tracker. We define thermalisation as the point at which the particle's kinetic energy is no longer atypical relative to the Maxwellian energy distribution of the background ions. Accordingly, the merge probability is taken to be the probability that an energy drawn from the thermal distribution exceeds $K_s^{(1,\text{lab})}$,
\begin{equation}
    P_{merge}=\min\left(1, \frac{\Gamma\left(\frac{3}{2},K_s^{(1,\text{lab})}\right)}{\Gamma\left(\frac{3}{2},\frac{3}{2}T_i\right)}\right),
\end{equation}
where $\Gamma(s,x)$ is the upper incomplete gamma function. The denominator normalises the expression such that all particles with $K_s^{(1,\text{lab})}\leq3T_i/2$ are thermalised.

\subsection{Discrete collisions}

A target particle from the background ion species must be selected for the discrete collision step. We independently sample candidates from each background ion species, and then choose one using a discrete distribution weighted by the instantaneous collision frequencies,
\begin{equation}
    \nu_{sb}=\Sigma_{sb} \sqrt{2K_{tot}^{(1,\text{cm})}\left(\frac{1}{m_s}+\frac{1}{m_b}\right)}.
\end{equation}
Here $\Sigma_{sb}=n_b\sum_j \sigma_{sb}^{(j)}$ is the total macroscopic cross-section and $K_{tot}^{(1,\text{cm})}$ is the total kinetic energy in the CoM frame. The collision channel is then chosen from a second discrete distribution weighted by the microscopic cross-sections $\sigma_{sb}^{(j)}$.

$K_{tot}^{(1,\text{cm})}$ can be computed from the kinetic energies of the incident and candidate target as well as their relative angle. The cosine of the relative angle is sampled from a uniform distribution $\mu_{sb}\sim U(-1, 1)$ because the plasma is isotropic. The energy of each candidate target is sampled from their respective Maxwellian distribution,
\begin{equation}
    K_b^{(1,\text{lab})}=T_b \gamma^{-1}\left(\frac{3}{2},\gamma'\sim U(0,1)\right),
\end{equation}
where $\gamma(s,x)$ is the lower incomplete gamma function. Since the selected candidate is knocked-on from the background plasma, we must also subtract $K_b^{(1,\text{lab})}$ from the kerma of the parent chain. The energies and direction vectors of the outgoing particles are then computed in the CoM frame and transformed back to the lab frame. For this we refer to the formalism in appendix \ref{appB}.

In the present implementation the available collisions channels are limited to elastic scattering and fusion reactions. This is in line with several other studies \cite{Wetering:PRE25l,Xue:SB25,Kumar87,Hirose:PPCF25}, although we recognise that inelastic scattering can make an appreciable contribution. For example for n-D scattering at $K_{tot}^{(1,\text{cm})}=10$ MeV the elastic and inelastic components are 0.6 b and 0.4 b, respectively \cite{ENDF:25}. Peres and Shvarts \cite{Peres75} showed that (n,2n) reactions can improve the chain-reaction multiplicity by several percent, although their spectra were poorly known at the time. Robinson \cite{Robinson24a} included endothermic D(p,n)2p disintegrations which acted as an energy sink for protons and delayed the onset of criticality by $\sim5$ keV. Therefore future work should include a more comprehensive list of inelastic scattering channels.

\subsection{Elastic scattering} \label{sec:elastic}

In an elastic scatter we keep tracking the suprathermal particle and also add the recoiling target to the tracker. From conservation of energy and momentum we have $K_{s/b}^{(1,\text{cm})}=K_{s/b}^{(2,\text{cm})}$ and $\hat{\mathbf{n}}_{s}=-\hat{\mathbf{n}}_b$, with $\hat{\mathbf{n}}_{s/b}$ the unit vectors for the outgoing particles' velocities. For the original particle this is written in spherical coordinates,
\begin{equation}
    \hat{\mathbf{n}}_{s}=
    \begin{bmatrix}
        \mu, & \sqrt{1-\mu^2}\cos\phi, & \sqrt{1-\mu^2}\sin\phi
    \end{bmatrix}.
\end{equation}

Elastic scattering is assumed to have azimuthal symmetry and so the azimuthal angle $\phi\sim U(0, 2\pi)$. By contrast, the cosine of the poloidal scattering angle has a distribution function $f(K_{tot}^{(1,\text{cm})},\mu)$, where $K_{tot}^{(1,\text{cm})}=K_s^{(1,\text{cm})}+K_b^{(1,\text{cm})}$. We choose $\mu$ by inverse transform sampling,
\begin{equation}
    \mu=F^{-1}\left(K_{tot}^{(1,\text{cm})},F'\sim U(0,1)\right), \label{eqn:ITS}
\end{equation}
where $F$ is the cumulative distribution function (CDF) from $\mu_m$ to $\mu$. The minimum value $\mu_m$ is $-1$ for distinguishable particles and 0 for identical particles. Often $F^{-1}$ is not analytic, so equation \ref{eqn:ITS} is actually a bilinear interpolation on a rectilinear grid $\left(K_{tot}^{(1,\text{cm})}, F'\right)\in \mathbb{R}_{\geq0}\times[0,1]$. The grid is refined so that the absolute uncertainty in $\mu$ is $<0.005$ when interpolating.

\begin{figure}
 \centering
        \includegraphics[width=\linewidth]{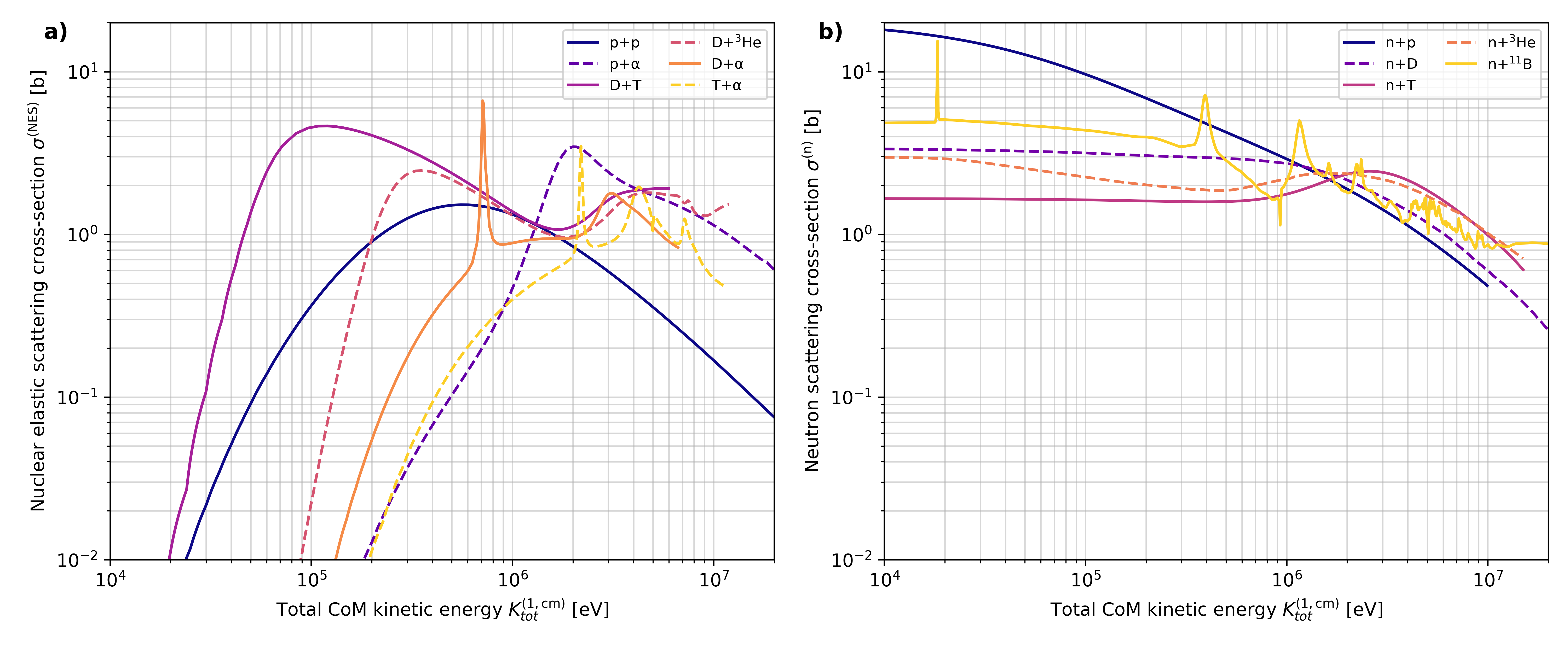}
 \caption{Cross-sections for (a) nuclear elastic scattering and (b) neutron elastic scattering. Data is extracted from the relevant ENDF/B-VIII.1 files \cite{ENDF:25}.}
\label{fig2}
\end{figure}

Angular distributions for neutron and charged particle scattering are taken from the open-access ENDF/B-VIII.1 file formats \cite{ENDF:25}. The cross-sections used in this work are shown in figure \ref{fig2}, which has been plotted with the same axis limits as figure \ref{fig1} to illustrate their relative likelihood.
In this paper we only present results for the NES term in the charged particle scattering distributions obtained from the ENDF files that have a nuclear amplitude expansion format. Other formats mix the pure nuclear and interference terms as a residual cross-section, which can have a negative angular distribution and therefore be nonphysical if the Coulomb term is not also considered. We save discussion of Monte-Carlo sampling of the Coulomb and interference terms for future work. Enhanced slowing down due to CLS is accounted for in the stopping model, and therefore any sampling scheme used to track the recoils must be consistent with this model.

\subsection{Two-body fusion reactions}

In a reaction we stop tracking the suprathermal particle and add the reaction products to the tracker. The two-body fusion reactions considered in this paper were presented in figure \ref{fig1}, except for p$^{11}$B which requires a three-body treatment. For the reaction products $r1$ and $r2$ the CoM frame kinetic energies are
\begin{equation}
    K_{r1/r2}^{(2,\text{cm})}=\frac{m_{r2/r1}}{m_{r1}+m_{r2}}K_{tot}^{(2,\text{cm})},
\end{equation}
where $K_{tot}^{(2,\text{cm})}=K_{tot}^{(1,\text{cm})}+Q$, with $Q$ the reaction energy. The procedure for selecting the unit vectors and transforming back to the lab frame is almost identical to that of elastic scattering, except we assume that all reactions are isotropic. Therefore $\hat{\mathbf{n}}_{r1}$ is chosen with $\mu\sim U(-1, 1)$ and $\phi\sim U(0,2\pi)$.

\subsection{p$^{11}$B fusion reactions}
The outgoing $\alpha$-particles from p$^{11}$B fusion have a continuous energy spectrum because it is a three-body breakup reaction. The breakup typically proceeds as a sequential decay via an intermediate $^8$Be nucleus, which is either in a ground ($\alpha_0$ channel) or excited state ($\alpha_1$ channel), the latter of which has a $\gtrsim 10$ times higher probability \cite{Sikora:JFE16}. Direct $3\alpha$ breakup is also possible but exceedingly rare. This work uses the spectrum model from Quebert and Marquez \cite{Quebert69}, which is described in detail in appendix \ref{appC}. 

We denote the CoM kinetic energy as $K_{\alpha1}^{(2,\text{cm})}$ for the first emitted $\alpha$-particle (primary) and as $K_{\alpha2}^{(2,\text{cm})}$ for the secondary. From conservation of energy and momentum it is simple to show that the kinematic region is bounded by an ellipse, $0\leq K_{\alpha1}^{(2,\text{cm})}\leq\frac{2}{3}K_{tot}^{(2,\text{cm})}$ and $\epsilon_{2,m}\leq K_{\alpha2}^{(2,\text{cm})}\leq\epsilon_{2,M}$, where
\begin{equation}
\epsilon_{2,m/M}=\frac{1}{2}\left|K_{tot}^{(2,\text{cm})}-K_{\alpha1}^{(2,\text{cm})}\pm\sqrt{K_{\alpha1}^{(2,\text{cm})}\left(2K_{tot}^{(2,\text{cm})}-3K_{\alpha1}^{(2,\text{cm})}\right)}\right|.
\end{equation}

We use a two-step inverse transform sampling process to determine the energies; first sample $K_{\alpha1}^{(2,\text{cm})}$ from the singles spectra $f$ and then sample $K_{\alpha2}^{(2,\text{cm})}$ from the coincidence spectra $g$, conditioned on $K_{\alpha1}^{(2,\text{cm})}$. $g$ is proportional to the transition matrix element and $f$ is its integral over $\epsilon_2$,
\begin{gather}
g\left(K_{tot}^{(1,\text{cm})},K_{\alpha1}^{(2,\text{cm})},K_{\alpha2}^{(2,\text{cm})}\right)\propto|\mathcal{M}_{if}|^2, \\
f\left(K_{tot}^{(1,\text{cm})},K_{\alpha1}^{(2,\text{cm})}\right)=\int_{\epsilon_{2,m}}^{\epsilon_{2,M}}g\left(\epsilon_2'\right)d\epsilon_2'. \label{eqn:singles}
\end{gather}
Inverse transform sampling then proceeds as follows
\begin{gather}
K_{\alpha1}^{(2,\text{cm})}=F^{-1}\left(K_{tot}^{(1,\text{cm})},F'\sim U(0,1)\right), \\
    K_{\alpha2}^{(2,\text{cm})}= G^{-1}\left(K_{tot}^{(1,\text{cm})},F',G'\sim U(0,1)\right),
\end{gather}
where the value of $F'$ must be reused when selecting $K_{\alpha2}^{(2,\text{cm})}$ to ensure it is conditioned on $K_{\alpha1}^{(2,\text{cm})}$. If we remove the conditioning requirement then a histogram of $K_{\alpha2}^{(2,\text{cm})}$ should yield a singles spectra identical to that of the primary since there is no way to distinguish between them at observation.

\begin{figure}[h]
 \centering
        \includegraphics[width=\linewidth]{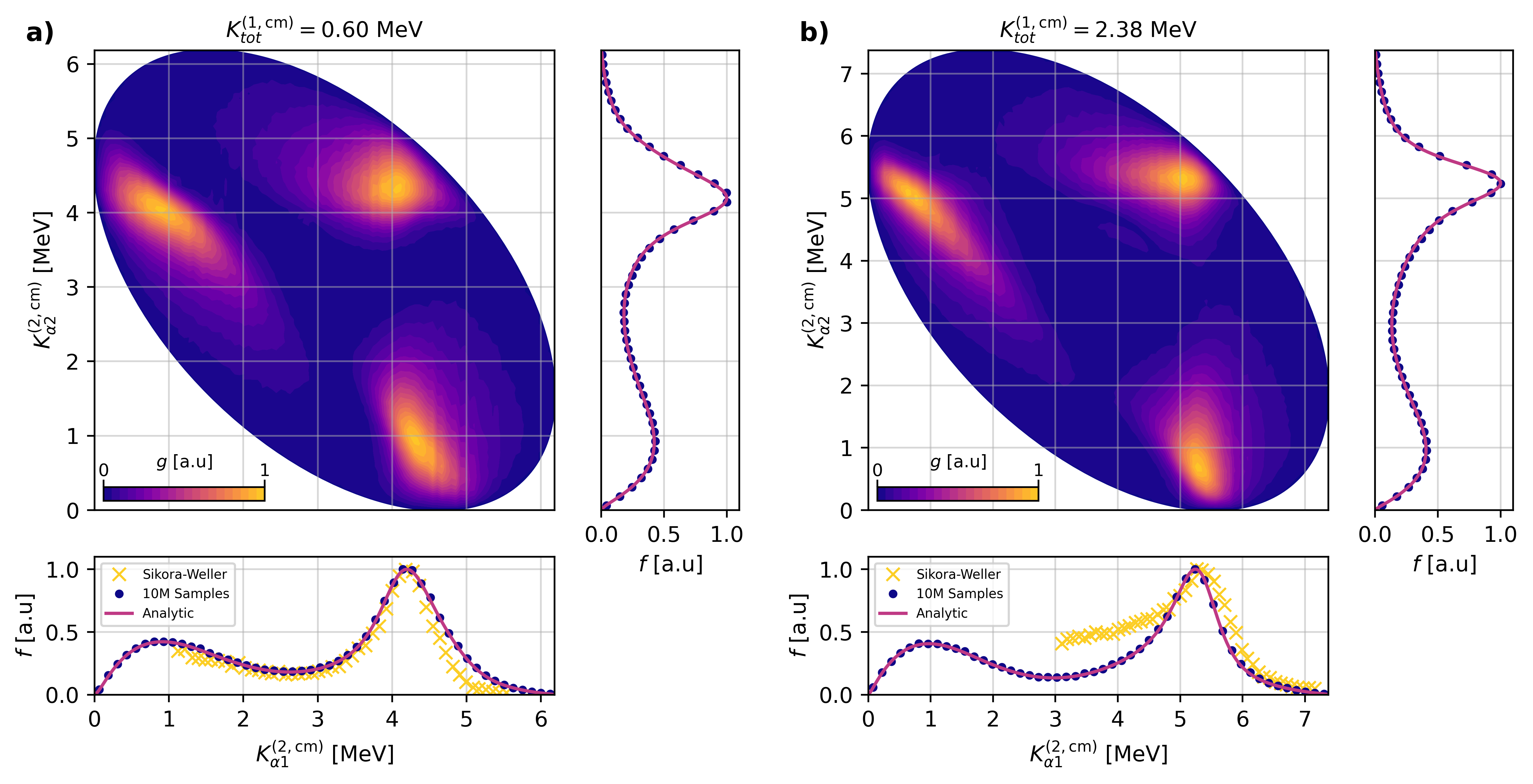}
 \caption{Results for inverse transform sampling of the Quebert and Marquez model \cite{Quebert69} at (a) $K_{tot}^{(1,\text{cm})}=600$ keV and (b) $K_{tot}^{(1,\text{cm})}=2.38$ MeV. The contour plot is the coincidence spectra and the line plots are the singles spectra for the primary and secondary $\alpha$-particle, which should be equal. The sampling histogram has been plotted alongside equation \ref{eqn:singles} and the observed spectra data from \cite{Sikora:JFE16}.}
\label{fig3}
\end{figure}

The inverse CDF's $F^{-1}$ and $G^{-1}$ are stored on rectilinear grids $\left(K_{tot}^{(1,\text{cm})},F'\right)\in\mathbb{R}_{\geq0}\times[0,1]$ and $\left(K_{tot}^{(1,\text{cm})},F',G'\right)\in\mathbb{R}_{\geq0}\times[0,1]^2$, respectively. $G^{-1}$ is deliberately mapped to $F'$ instead of $K_{\alpha1}^{(2,\text{cm})}$ so that interpolation takes advantage of the regular grid structure and avoids expensive triangulation. The grid for $F^{-1}$ is refined so that the absolute uncertainty in $K_{\alpha1}^{(1,\text{cm})}$ is $<10$ keV during interpolation. The grid for $G^{-1}$ has an additional dimension and so memory constraints limited the interpolation uncertainty for $K_{\alpha2}^{(1,\text{cm})}$ to $<100$ keV. Despite this, the singles spectra for the primary and secondary are still identical and converge to the analytic model, as shown figure \ref{fig3}.

Also plotted is the experimental data that was used to re-evaluate the p$^{11}$B cross-section \cite{Sikora:JFE16}, which have been normalised to arbitrary units. A key limitation of our analytic model is that it can only be parameterised for a single channel, which we take to be the $\alpha_1$ decay at the dominant $K_{tot}^{(1,\text{cm})}=620$ keV resonance. The model matches the experimental data for $K_{tot}^{(1,\text{cm})}=600$ keV very well, save for a slight overestimation at high energy possibly due to lack of the $\alpha_0$ channel. However the model does diverge for $K_{tot}^{(1,\text{cm})}=2.38$ MeV since the reaction proceeds through different angular momentum states ($J=3^-$, $\ell=1$). At these energies the cross-section is $\sim50$\% of the peak, so the effects of a softer $\alpha$-spectrum may still be considerable. Although the use of nuclear physics-informed spectrum is an improvement over previous studies \cite{Belloni:PP18,Hirose:PPCF25}, future work should look to develop a spectrum model that can account for different resonances.

The emission direction of the primary $\alpha$-particle energies is isotropic with $\mu_{\alpha1}\sim U(-1,1)$ and $\phi_{\alpha1}\sim U(0,2\pi)$. The angle between the primary and secondary is then fixed by energy and momentum conservation
\begin{equation}
\mu_{12}=\hat{\mathbf{n}}_{\alpha1}\cdot\hat{\mathbf{n}}_{\alpha2}=\frac{\frac{1}{2}K_{tot}^{(2,\text{cm})}-K_{\alpha1}^{(2,\text{cm})}-K_{\alpha2}^{(2,\text{cm})}}{\sqrt{K_{\alpha1}^{(2,\text{cm})}K_{\alpha2}^{(2,\text{cm})}}}.
\end{equation}
Due to the isotropic breakup of the $^8$Be nucleus, $\hat{\mathbf{n}}_{\alpha2}$ is also rotated azimuthally around $\hat{\mathbf{n}}_{\alpha1}$ by $\phi_{12}\sim U(0,2\pi)$. The poloidal and azimuthal rotations therefore combine to the following expression
\begin{equation}
    \hat{\mathbf{n}}_{\alpha2}=\mu_{12}\hat{\mathbf{n}}_{\alpha1} + \sqrt{1-\mu_{12}^2}\left(\cos\phi_{12}\hat{\mathbf{e}}+\sin\phi_{12}\hat{\mathbf{n}}_{\alpha1}\times\hat{\mathbf{e}} \right),
\end{equation}
where $\hat{\mathbf{e}}$ is an axis perpendicular to $\hat{\mathbf{n}}_{\alpha1}$. A simple choice is
\begin{equation}
    \hat{\mathbf{e}}=
    \begin{bmatrix}
        -\sqrt{1-\mu_{\alpha1}^2}, & \mu_{\alpha1}\cos\phi_{\alpha1}, & \mu_{\alpha1}\sin\phi_{\alpha1}
    \end{bmatrix}.
\end{equation}
Once the primary and secondary $\alpha$-particle energies have been determined in the lab frame, the lab-frame energy of the remaining $\alpha$-particle is simply given by energy conservation.

\section{Results}
We divide this section broadly into four parts. First, we compare our model to the results from Robinson \cite{Robinson24a}. Second, we revisit Moreau's \cite{Moreau:NF77} evaluation of the energy multiplication for fast protons in `cold' boron plasma. We then extend this analysis to the case of fast protons, $\alpha$-particles and neutrons in ICF-relevant borane ($^{11}$BH$_3$) plasmas. Finally, we compute the suprathermal energy gain for reaction-rate-weighted fusion product spectra for DT, pure deuterium, $^{11}$BH$_3$ and enriched borane $^{11}$BHDT.

\subsection{Validation with pure deuterium} \label{sec:DinD}

\begin{figure}[b]
 \centering
        \includegraphics[width=\linewidth]{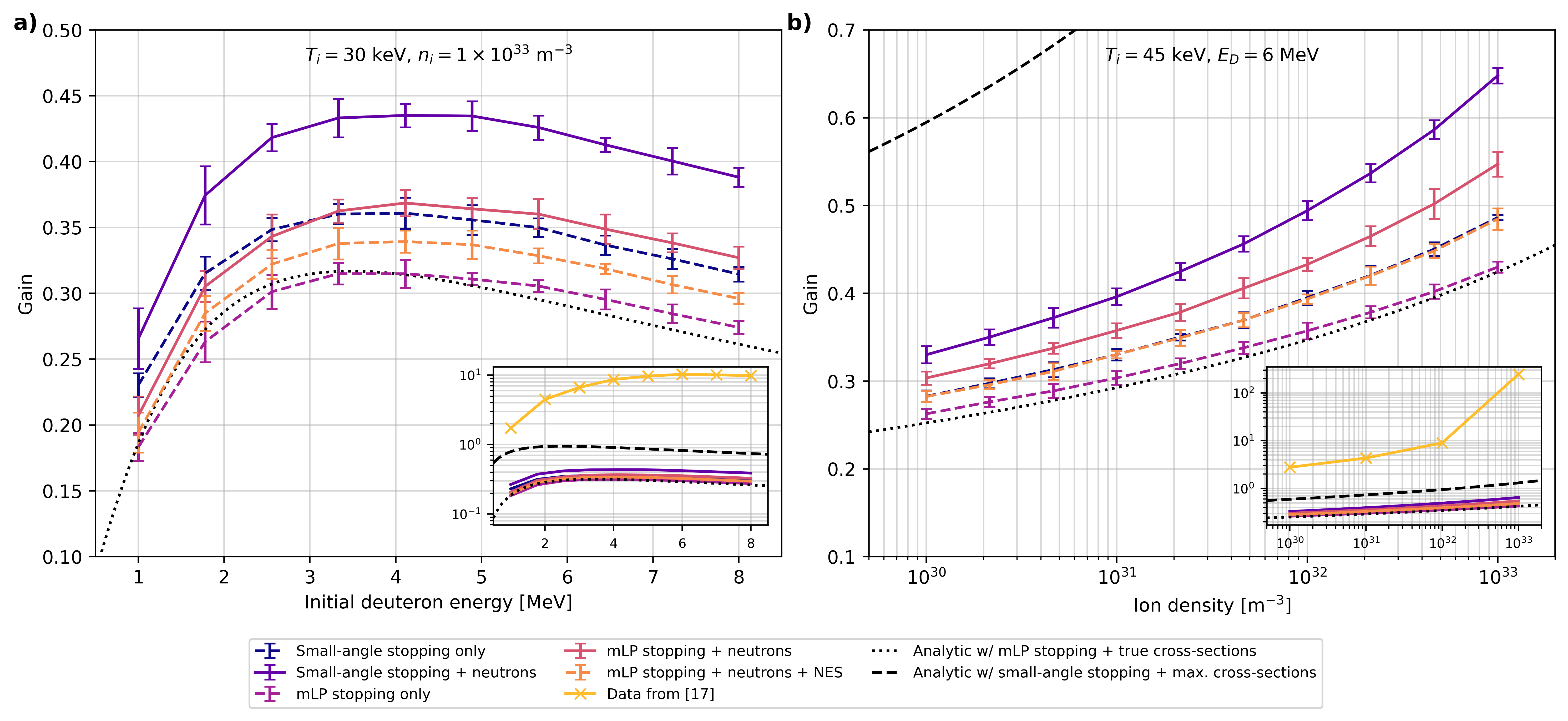}
 \caption{Dependence of gain on (a) deuteron energy and (b) ion number density for fast deuterons in deuterium, corresponding to figure 6 and figure 8 in \cite{Robinson24a}. The published results from Robinson are plotted in the inset axes. Error bars correspond to $\pm 3\sigma$ after 10 simulation batches with $10^5$ particles each.}
\label{fig4}
\end{figure}

The dependence of the suprathermal energy multiplication gain for fast deuterons in deuterium as a function of the initial energy $E_D$ (1--8 MeV) and target ion density (10$^{30}$--10$^{33}$ m$^{-3}$) has been studied. This system is of interest for direct comparison with Robinson's work \cite{Robinson24a} using a more complete physics model that includes updated stopping powers, realistic thermal broadening, and nuclear elastic scattering (NES). Figure \ref{fig4} compares the gain for five different physics models:
\begin{enumerate}
    \item Use of a small-angle electronic and ionic stopping power model \cite{Atzeni:book04} with neutron and NES scattering not included. This stopping model is equivalent to removing the dielectric and $1/\ln\Lambda$ CLS terms in equation \ref{eqn:mLP}. Here only in-flight reactions from the primary deuteron and triton and helion burn products contribute to the gain.
    \item Same as (1) except with deuteron up-scattering by neutrons included. This is broadly equivalent to Robinson's model \cite{Robinson24a}, save for the fact that we do not include endothermic D(p,n)2p disintegrations. %As a result our results should be slightly more optimistic.
    \item Same as (1) except with the modified Li-Petrasso (mLP) stopping model \cite{Zylstra:PP19}. This is the most pessimistic model due to reduced in-flight reaction probability and the absence of knock-ons.
    \item Same as (3) except with neutron scattering included.
    \item Same as (4) except with NES included, which in this case is T-D, $^3$He-D and $\alpha$-D scattering. Notably absent is D-D scattering which is saved for future work since its corresponding ENDF file cannot be sampled without a CLS model, as explained in section \ref{sec:elastic}.
\end{enumerate}

In all cases the gain at 30 keV has a broad maximum for 3--5 MeV deuterons. For 6 MeV deuterons, increasing the temperature to 45 keV improves the gain by $50$\%, and there is a weak dependence on density considering that it spans three orders of magnitude. For the highest observed gain at $T_i=45$ keV, $E_D=6$ MeV and $n_i=10^{33}$ m$^{-3}$, the mLP stopping power reduces $G$ by 15\%, neutron scattering increases $G$ by 30\% and, compared to case (4), NES reduces $G$ by 10\%. In fact, the combined reductions from mLP stopping and NES result in gains that are comparable to that of case (1). 

Also plotted in the inset axes is data from Robinson's Monte Carlo code \cite{Robinson24a}, which differ from our results by over an order of magnitude in some cases. For benchmarking we provide a simple analytic model that only requires the fusion cross sections and stopping power, both of which have been validated. A fast deuteron has an in-flight fusion probability $p_1$ during slowing down, upon which it has roughly equal chance of producing a $3$ MeV proton and $1$ MeV triton (with its own in-flight fusion probability $p_2$), or a $2.5$ MeV neutron and $0.8$ MeV helion (also with its own in-flight fusion probability $p_3$). These energies are fixed and we ignore the kinematic boosting of reaction products expected at higher collision energies. Fast neutrons will successively transfer 4/9ths of their energy per n-D scatter event, and the total number of DD fusions from all the scatters is the sum of each recoil's in-flight fusion probability. We denote these $k_2$ and $k_3$ for $14.1$ MeV DT and $2.5$ MeV DD neutrons, respectively. The total energy released by a DD fusion event can thus be found recursively
\begin{equation}
    Q_\text{DD}=\frac{1}{2}\left(Q_\text{D(d,p)} + p_2(Q_\text{T(d,n)}+k_2Q_\text{DD})\right) + \frac{1}{2}\left(Q_\text{D(d,n)} + p_3Q_{^3\text{He(d,p)}}+k_3Q_\text{DD}\right),
\end{equation}
and therefore the gain for initial energy $E_D$ is
\begin{equation}
G\approx\frac{p_1\left(Q_\text{D(d,p)}+Q_\text{D(d,n)}+p_2Q_\text{T(d,n)}+p_3 Q_{^3\text{He(d,p)}}\right)}{\left(2-p_2k_2-k_3\right)E_D}.
\end{equation}
Each in-flight fusion probability is equal to $1-\exp(-\Delta N)$, where $\Delta N$ from equation \ref{eqn:deltaN} is computed with $\bar{\nu}_{sb}=\Sigma_{sb} v_s$ to remove the effect of cross-section thermal broadening. The extent to which the analytic model underestimates the gain of case (4) increases from 10\% at 1 MeV to 20\% at 8 MeV. Also plotted is the same analytic model with small-angle stopping and cross-sections fixed at their maximum values (e.g. 5 barns for DT) as an upper bound on the results.

\subsection{Validation with beam-driven p$^{11}$B}
\begin{figure}[h]
 \centering
        \includegraphics[width=0.5\linewidth]{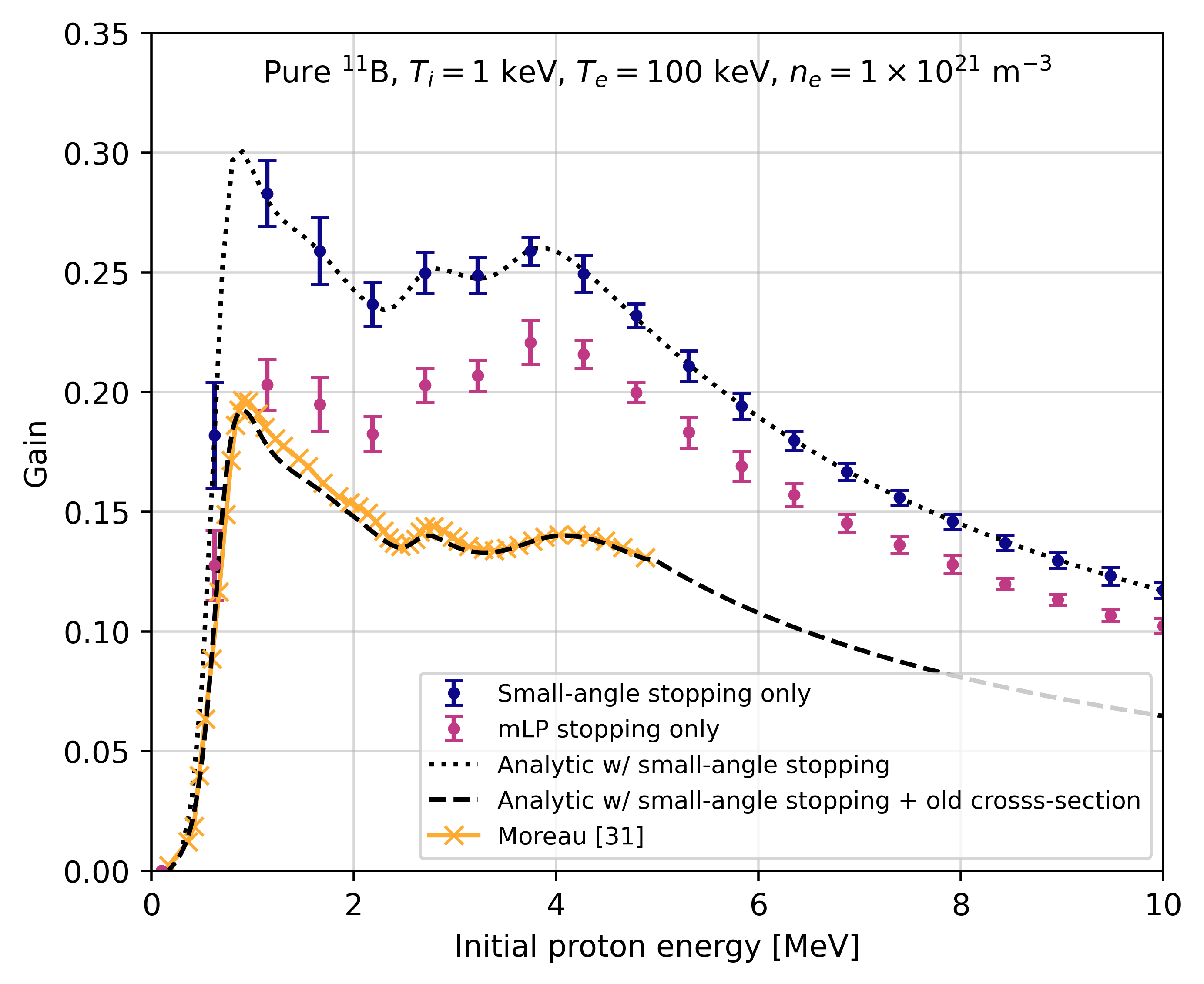}
 \caption{Revision of figure 6 in \cite{Moreau:NF77} with mLP stopping power and latest p$^{11}$B cross-section. Error bars correspond to $\pm 3\sigma$ after 10 simulation batches with $10^5$ particles each.}
\label{fig5}
\end{figure}

Calculation of the energy multiplication factor for fast protons in $^{11}$B was carried out by Moreau \cite{Moreau:NF77} for a hot electron plasma ($T_i$=1 keV, $T_e>$50 keV) at $n_e=10^{21}$ m$^{-3}$, a regime relevant for magnetic fusion schemes. The small-angle stopping model \cite{Atzeni:book04} was used alongside an outdated cross-section which underestimates the current p$^{11}$B cross-section by at least 30\% \cite{Tentori:NF23}. Figure \ref{fig5} plots the gain using the analytic in-flight fusion probability for small-angle stopping as described in the previous section. The Monte Carlo model matches the analytic estimate since thermal broadening at such low ion temperatures is negligible. For validation we use the same model with the old cross-section data to replicate figure 6 in \cite{Moreau:NF77}. Including mLP stopping with the new cross-section shifts the peak to 4 MeV with a maximum value of 0.22---only a small improvement to Moreau's original results.

\subsection{Fast particles in $^{11}$BH$_3$} \label{sec:fastpB11}
\begin{figure}
 \centering
        \includegraphics[width=\linewidth]{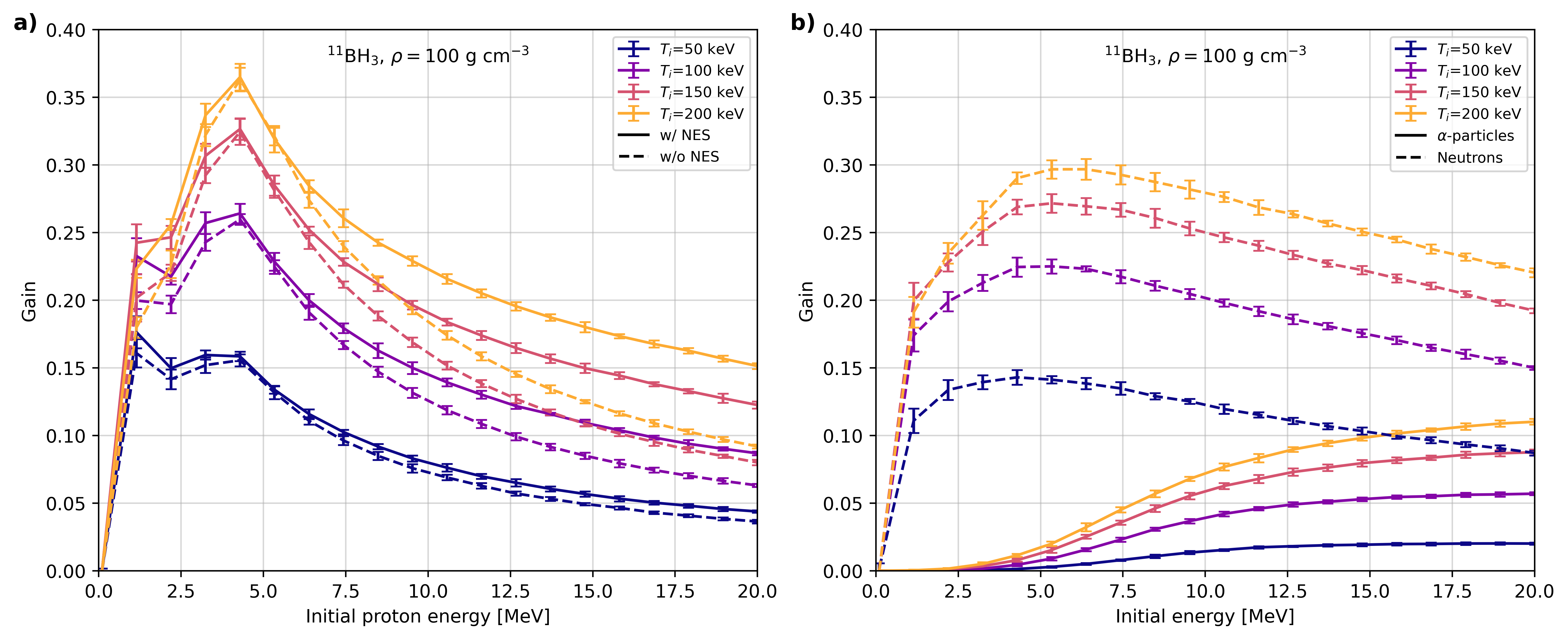}
 \caption{Suprathermal energy gain for (a) protons with and without NES and (b) $\alpha$-particles and neutrons with NES in high-density $^{11}$BH$_3$. Error bars correspond to $\pm 3\sigma$ after 10 simulation batches with $10^5$ particles each.}
\label{fig6}
\end{figure}

The case of fast protons in isothermal $^{11}$BH$_3$ at ICF conditions has been investigated for its applicability to a proton fast-ignition scheme \cite{Mehlhorn:LPB22}, and results with and without NES at 100 g cm$^{-3}$ are plotted in figure \ref{fig6}a. The optimum proton energy is still 4 MeV despite ten orders of magnitude difference in density from the previous section. The peak gain exhibits a strong temperature dependence, though the rate of increase progressively weakens at higher $T_i$, so gains greater than 0.4 may not actually be possible. The gain improves when p-p NES is included, although its impact is only significant at higher proton energies and does not appreciably modify the peak.

Figure \ref{fig6}b plots the gain for fast $\alpha$-particles and neutrons with NES included. Again we see diminishing returns when increasing the temperature, and from the trend it appears that gains in excess of 0.15 for $\alpha$-particles and 0.35 for neutrons may not be possible. $\alpha$-particles in particular exhibit quite a poor suprathermal enhancement especially for energies typically observed in p$^{11}$B spectra ($<9$ MeV). The gain from fast neutrons, on the other hand, is comparable to that of fast protons despite not having an in-flight reaction.

\begin{figure}[!b]
 \centering
        \includegraphics[width=\linewidth]{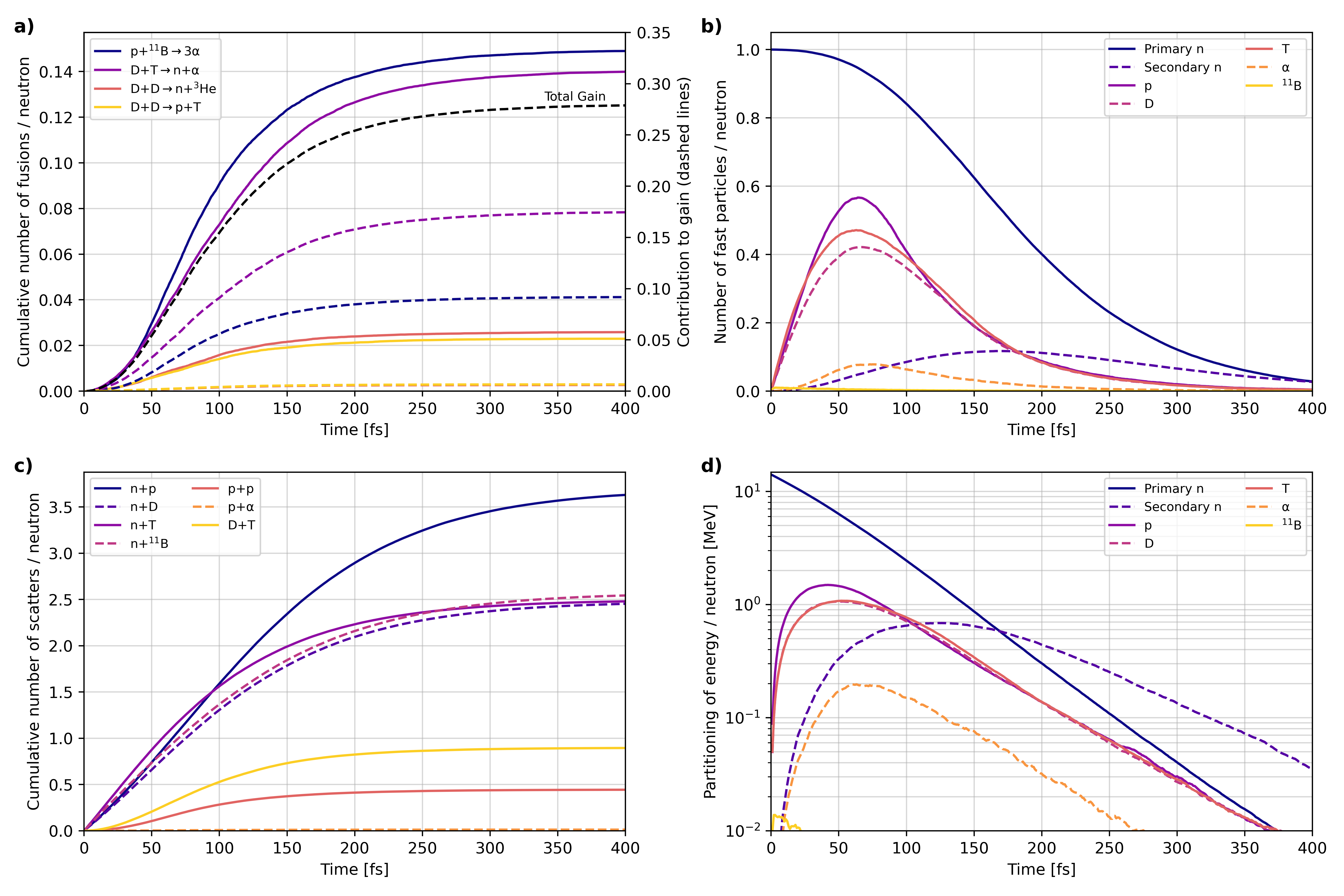}
 \caption{Time-dependent results for a 14.1 MeV neutron in $^{11}$BHDT at $T_i=50$ keV and $\rho=100$ g cm$^{-3}$, averaged over $10^5$ simulations. (a) The time-integrated number of fusion events. (b) The number of particles for each species at any given moment, distinguishing between primary neutrons and secondary neutrons produced from fusion. (c) The time-integrated number of neutron and NES scatters. (d) The average partitioning of energy into each species.}
\label{fig7}
\end{figure}

If the generation of such energetic neutrons were to come from DT fusion, then a natural extension is to investigate the suprathermal enhancement in enriched borane, $^{11}$BHDT. Figure \ref{fig7} is a time-dependent study of 14.1 MeV neutrons interacting with a $T_i=50$ keV, $\rho=100$ g cm$^{-3}$ $^{11}$BHDT plasma. The total gain is 0.28, approximately three times the value for $^{11}$BH$_3$ at the same condition. p$^{11}$B only contributes 32\% of the gain even though there are slightly more p$^{11}$B fusions than DT fusions on average. The peak fusion rate, which is given by the inflexion point in the total gain curve in figure \ref{fig7}a, occurs at 70 fs. This coincides with the peak number of protons, deuterons, tritons and $\alpha$-particles in figure \ref{fig7}b. For every initial neutron, 0.57 protons, 0.42 deuterons and 0.47 tritons will exist at this point. By contrast in figure \ref{fig7}c neutron scattering produces slightly more tritons than protons, although the protons will take longer to thermalise because they have on average 33\% more energy (n-p scattering transfers 1/2 the neutron energy vs 3/8 for n-T scattering). This is consistent with the peak energies in figure \ref{fig7}d. The stopping time of neutrons is longer still, delaying the peak in the average number and total energy for the secondary neutrons.

As a final observation, we note that even though more than half of the fusion events are p$^{11}$B, the number of subsequent p-$\alpha$ scattering events is negligible. The suprathermal chain is instead driven by primary neutron scatters with mild contributions from secondary D-T and p-p scatters.

\subsection{Suprathermal energy gain in thermonuclear fuels}
\begin{figure}[!b]
 \centering
        \includegraphics[width=\linewidth]{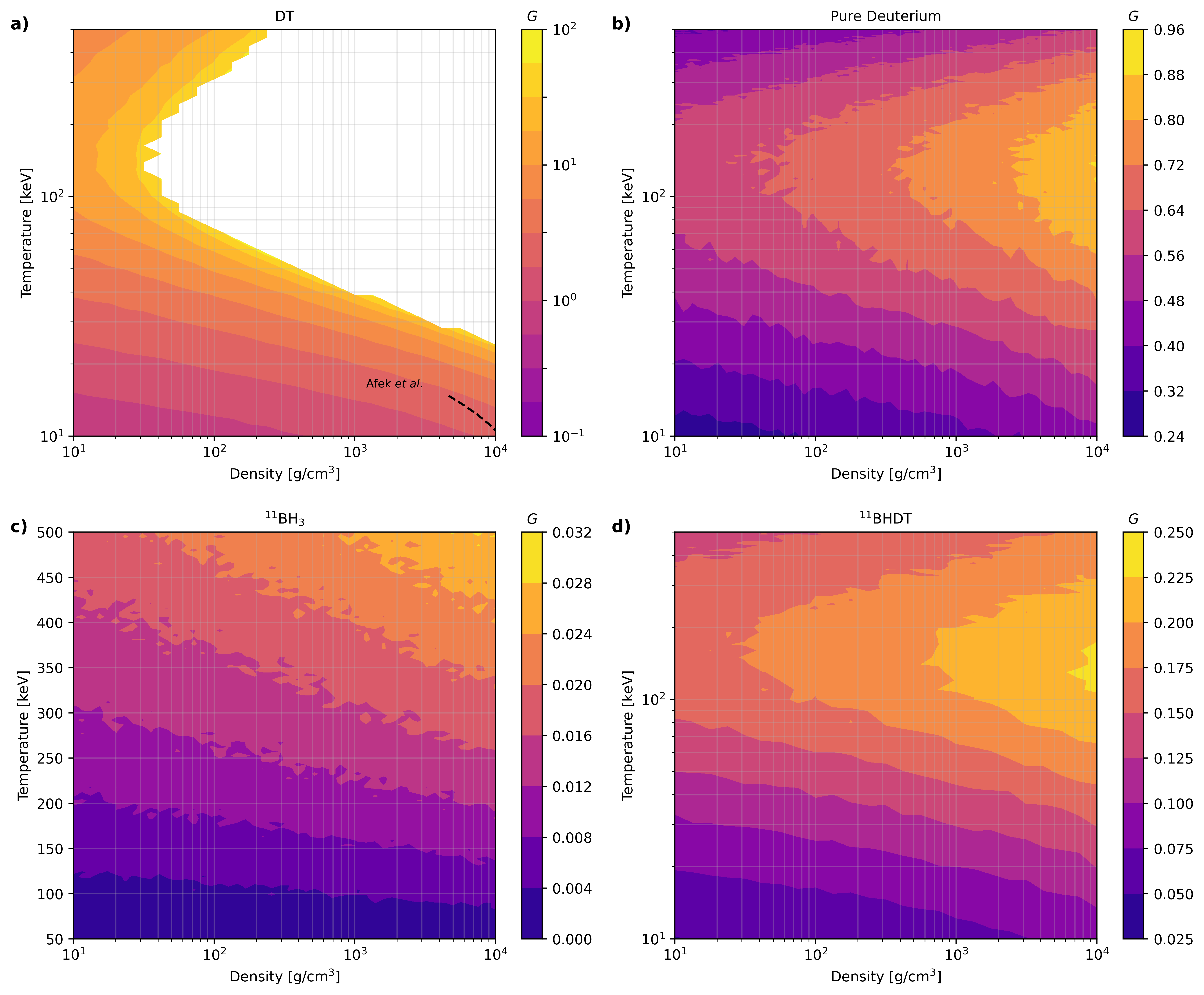}
 \caption{Suprathermal energy gains as a function of density and temperature for (a) DT, (b) pure deuterium, (c) $^{11}$BH$_3$ and (d) $^{11}$BHDT. Each point is the mean of five simulation batches with $10^4$ particles each. The relative uncertainty for $\pm1\sigma$ is less than 10\%. In (a) we also plot the criticality threshold from \cite{Afek:JPD78} for DT with an undisclosed amount of $^6$Li.}
\label{fig8}
\end{figure}

The suprathermal energy gain per average reaction in the $\rho$--$T_i$ parameter space has been plotted for DT, deuterium, $^{11}$BH$_3$ and $^{11}$BHDT in figure \ref{fig8}. Average reaction refers to initialising the incident particles by weighted sampling of each possible fusion reaction channel, using the reaction rate $n_1n_2\langle\sigma v\rangle$ \cite{Atzeni:book04} as the weights. We limit our parameter space to $10\leq\rho\leq10^4$ g cm$^{-3}$ and $10\leq T_i\leq500$ keV to keep results relevant to present and next-generation laser driver capabilities.

DT can transition to super-criticality and for comparison the critical threshold from Afek \textit{et al.} \cite{Afek:JPD78} is plotted in figure \ref{fig8}a. Their results were obtained using a multi-species kinetic transport model analogous to neutron-transport equations used for studying fission systems \cite{Peres75}. For the little overlap in the parameter space our model estimates criticality at approximately double the temperature. Their work, however, used the small-angle stopping model, relied on the sparse data for NES that was available at the time and, most crucially, only considered DT with an undisclosed amount of $^6$Li for tritium breeding.

The gains in all four contour plots demonstrate a strong dependence on temperature and a much weaker dependence on density. For the fuels with deuterium an optimum ion temperature of 100--200 keV maximises the gain for a given density. For pure $^{11}$BH$_3$ the maximum enhancement observed is on the order of a few percent, which is in line with past analytic results \cite{Belloni:LPB22, Hirose:PPCF25}. An optimum ion temperature may exist, although at far more difficult conditions than was is surveyed.

\section{Discussion}

Our results show that deuterium cannot become critical, even when taking the fusion cross-sections at their maximum value. Neutron-driven deuteron up-scattering makes a meaningful contribution to the gain, although this is partially mitigated by the enhanced charged particle stopping in the modified Li–Petrasso stopping model. A further reduction in the gain is observed with the inclusion of D-T, D-$^3$He and D-$\alpha$ NES. A time-dependent analysis of the results with NES has showed that secondary neutrons from DT fusions have a lower average energy due to a `softening' of the triton slowing down spectrum at higher energies. These neutrons then produce deuterons with less energy and lower in-flight reaction probability. We however do not account for D-D scattering. Primary knock-ons will likely contribute a net increase to the gain, as shown for protons in p$^{11}$B, although its effect will be on the order of 10\% and not significantly change our conclusions.

In-flight and knock-on fusions provide at most an additional 40\% of the energy deposited in a p$^{11}$B proton fast-ignition scheme, optimised at 4 MeV injection energy. Following Moreau \cite{Moreau:NF77}, a gain of 40\% requires an energy conversion efficiency of 71\% to generate net energy from beam-target fusions, which is unreasonable even before factoring in the beam injection efficiency and losses used to create and sustain the plasma. The suprathermal enhancement improves with temperature and so these effects cannot be immediately exploited when heating cold fuels.

At low temperatures the ion range and fusion probability is reduced due to enhanced dielectric stopping. In figure \ref{fig8} the enhancement also decreases in the high temperature limit, which is $>100$--200 keV for fuels that rely on DT fusion. In this regime the average thermal energy per particle is well in excess of the peak of the cross-section. The weak positive dependence of the suprathermal energy gain on density is evident in equation \ref{eqn:deltaN}; the factor of $n_i$ in both the collision frequency and stopping power cancel, so that density only enters via the Debye length $\lambda_D\propto\sqrt{T_i/\rho}$ in the Coulomb logarithm. We note that for $\rho>10^4$ g cm$^{-3}$ the effects of electron degeneracy become significant since the effective electron temperature grows like $T_e\propto\rho^{2/3}$. Son and Fisch \cite{Fisch:PLA04} show that the stopping power no longer has a density dependence in this regime, and thus equation \ref{eqn:deltaN} will be proportional to the density. Consequently the suprathermal gain will increase significantly and may unlock critical regimes even for pure p$^{11}$B.

Of the fuels studied, DT is likely to exhibit suprathermal effects at conditions relevant to present and next-generation driver capabilities, assuming that neutrons can couple to ions. The criticality threshold scales like $T_i\sim 200\rho^{-0.23}$, with $T$ in keV and $\rho$ in g cm$^{-3}$. The scattering cross section for 14.1 MeV neutrons in a 50/50 DT mix is 0.8 barns, and so the mass attenuation length is $\rho\lambda=5.2$ g cm$^{-2}$. If we assume the hotspot size $\rho R_{HS}$ is five times this then the scaling for energy is
\begin{gather}
    E_{HS}=4\pi r^3 n_iT_i\sim1.7\times10^9\rho^{-2.23},
\end{gather}
with $E_{HS}$ in MJ and $\rho$ in g cm$^{-3}$. At the extreme case of $\rho=10^4$ g cm$^{-3}$ then $R_{HS}=26$ $\mu$m and $E_{HS}\approx2$ MJ. For comparison, the first achievement of scientific breakeven on NIF had $\rho R_{HS}=0.44$ g cm$^{-3}$, $R_{HS}=67.5$ $\mu$m and $E_{HS}=118$ kJ \cite{AbuShawareb:PRL24}. A more detailed analysis of neutron stopping in finite in targets is clearly needed, although this simple result suggests that suprathermal enhancement from neutron knock-ons may have a significant contribution if such high density and $\rho R_{HS}$ regimes can be achieved.

Although it is non-critical, mixing DT with $^{11}$BH$_3$ yields a ten times greater enhancement than $^{11}$BH$_3$ alone due to the efficient energy transfer from neutrons to protons. In doing so the protons also act as a sink for the neutron energy which suppresses the DT chain reaction. Despite this, BHDT is a non-cryogenic fuel and may simplify target engineering requirements if a feasible burn-space is shown to exist, which we leave for future work. For pure $^{11}$BH$_3$ we corroborate with recent results \cite{Belloni:PPCF21, Hirose:PPCF25} that reject the premise of an $\alpha$-p avalanche mechanism. The charged particle stopping power scales like $\propto m_sZ_s^2$ for the same energy, so the drag on $\alpha$-particles is 16 times higher than for protons which makes them a poor intermediary of suprathermal chains. Moreover, our results assume isothermal ions and electrons, whereas ignition criteria require $T_e/T_i<0.5$ \cite{Ghorbanpour:FP24}, which will further increase the stopping power. We do not believe that further modification of the p$^{11}$B $\alpha$-spectrum, such as including the $\alpha_0$-channel or other resonances, will significantly change our results. This is supported by figure \ref{fig7}b which shows that the gain for typical 4 MeV $\alpha$-particles is at most 2\%, and even for 10 MeV $\alpha$-particles it does not exceed 7\%.

To conclude this section we identify opportunities for further exploration:
\begin{itemize}
    \item CLS and the CLS-NES interference in the scattering kernel, with additional considerations for identical particles and spin coupling \cite{ENDF:25}. Any scattering model \cite{Xue:SB25, Turrell:PRL14, Angus:JCR25} will need to replicate the $\sim1/\ln\Lambda$ large-angle correction in the stopping power. We note that the Coulomb section scales like $\propto 1/K_s^2$ and so its efficacy for enabling high-energy suprathermal chains may be limited.
    \item A more comprehensive list of scattering processes, for example (n,2n), (p,n) and neutron inelastic channels. In particular (n,2n) reactions were found to improve the DT chain reaction multiplicity by several percent \cite{Peres75}. Inclusion of neutron inelastic scattering may require a more thorough code which also tracks nuclear isomers, which may not undergo the same reactions as their ground states.
    \item Extension of the collision physics to relativistic regimes, which is particularly relevant for p$^{11}$B \cite{Lavell24}. 
    \item $^6$Li and $^7$Li seeding to explore low-tritium fuels and/or tritium breeding in situ \cite{Afek:JPD78}.
    \item The effect of finite geometries and more complete burn physics (e.g. fusion reactivities, bremsstrahlung, ion-electron equilibration) on the suprathermal enhancement with focus on target designs that aim to trap and/or multiply neutrons.
    \item Distortion of the background ion distributions away from the Maxwellian due to ion slowing down, which can enhance the reactivity. Note that this is a different suprathermal effect to the fast fusion chain reactions considered in this paper. In particular, the FOKN code by Zimmerman \textit{et al.} \cite{Zimmerman76} should be revisited with updated cross-sections since it provides the most comprehensive 0D formalism for studying kinetic effects.
\end{itemize}

\section{Conclusions}

We have developed a detailed Monte-Carlo model to evaluate the suprathermal enhancement in DT, deuterium, and borane fusion fuels, incorporating updated stopping powers, full thermal broadening, and nuclear elastic scattering. The results provide a clarified view of suprathermal chain processes under inertial fusion conditions and their influence is far smaller for advanced fuels than previously suggested.

We have not found evidence of a self-sustaining deuterium chain reaction for $\rho<10^4$ g cm$^{-3}$, $T_i<500$ keV. In $^{11}$BH$_3$, fast protons can produce a gain of at most 40\%, which may assist with the heating in a proton fast-ignition scheme. The enhancement increases at higher temperatures and we see a robust optimum at 4 MeV injection energy. $\alpha$-p avalanche mechanisms are ruled out due to large Coulomb drag, and for thermal $^{11}$BH$_3$ the enhancement is a few percent. Mixed $^{11}$BHDT fuel benefits from neutron-driven up-scattering, producing larger gains than for pure $^{11}$BH$_3$ systems, though still below criticality. If neutrons can be made to couple to ions, then DT may exhibit a fusion chain reaction at very high temperatures and densities. Even small amounts of neutron trapping is likely to boost the yield at plasma regimes accessible in the near term and should be factored into burn models.

The framework developed here establishes a foundation for future studies to include additional reactions and
scattering processes and more realistic capsule physics. We emphasise that non-Maxwellian fuel ion distributions
from burn product slowing down should still be revisited with a kinetic model that accounts for multiple reaction
channels and Coulomb and nuclear elastic scattering. While the present study constrains suprathermal gains in
conventional configurations, the vast, unexplored parameter space of advanced fuels leaves open the possibility of
entirely new target designs.

\begin{acknowledgments}
This work was supported by a collaboration between HB11 Energy and The University of New South Wales Sydney. MB would like to thank Max Tabak and Esmat Ghorbanpour for their sustained engagement and insightful discussions during the development of this work.
\end{acknowledgments}

\appendix

\section{Derivation of collision frequency} \label{appA}
The Maxwellian-averaged collision frequency between a fast particle with velocity $\mathbf{v}_s$ and background particle with velocity $\mathbf{v}_b$ is
\begin{equation}
    \bar{\nu}_{sb}=\left(\frac{a}{\pi}\right)^{3/2}\int_{\mathbb{R}^3}\Sigma_{sb}u_{sb} e^{-av_b^2}d^3\mathbf{v}_b,
\end{equation}
where $u_{sb}=|\mathbf{v}_s-\mathbf{v}_b|$ and $a=m_b/2T_b$. We can rewrite in spherical coordinates with polar axis along $\mathbf{v}_s$,
\begin{equation}
    \bar{\nu}_{sb}=\frac{2a^{3/2}}{\sqrt{\pi}} \int_0^\infty v_b^2e^{-av_b^2}\int_{-1}^1 \Sigma_{sb}u_{sb}d\mu_{sb} dv_b.
\end{equation}
Since $u_{sb}=\sqrt{v_s^2+v_b^2-2v_sv_b\mu_{sb}}$, the differential $d\mu_{sb}=-(u_{sb}/v_sv_b)du_{sb}$ and $\mu_{sb}\in[-1,1]$ maps to $u_{sb}\in[|v_s-v_b|,v_s+v_b]$. Hence
\begin{equation}
    \bar{\nu}_{sb}=\frac{1}{v_s}\frac{2a^{3/2}}{\sqrt{\pi}}\int_0^\infty v_be^{-av_b^2}\int_{|v_s-v_b|}^{v_s+v_b} \Sigma_{sb}u_{sb}^2du_{sb} dv_b.
\end{equation}
We can swap the order of integration since for $u_{sb}>0$, $v_b$ runs from $|v_s-u_{sb}|$ to $v_s+u_{sb}$,
\begin{equation}
    \bar{\nu}_{sb}=\frac{1}{v_s}\frac{2a^{3/2}}{\sqrt{\pi}}\int_0^\infty \Sigma_{sb}u_{sb}^2 \int_{|v_s-u_{sb}|}^{v_s+u_{sb}} v_be^{-av_b^2} dv_b du_{sb},
\end{equation}
leading to the final closed form
\begin{equation}
    \bar{\nu}_{sb}=\frac{1}{v_s}\sqrt{\frac{a}{\pi}}\int_0^\infty \Sigma_{sb}u_{sb}^2 \left[e^{-a(v_s-u_{sb})^2}-e^{-a(v_s+u_{sb})^2}\right]du_{sb}.
\end{equation}

For NES and neutron scattering the integral limits are taken from the available ENDF/B-VIII.1 data range. Although results with CLS are not presented in this paper, we note that the above integral does not converge for a $\sim1/u_{sb}^4$ Coulomb cross-section. In this case we instead encourage use of the slowing down relaxation rate derived from the Fokker-Planck equation \cite{Li93}. 

\section{Transforming between collision frames} \label{appB}
In a collision between a fast particle and background particle we define $\mathbf{v}_{s/b}$ and $\mathbf{u}_{s/b}$ as the initial velocities in the lab and CoM frame, respectively. The CoM velocity is $\mathbf{v}_c=(m_s\mathbf{v}_{s}+m_b\mathbf{v}_{b})/M$, for total mass $M=m_s+m_b$, and so $\mathbf{u}_{s/b}=\mathbf{v}_{s/b}-\mathbf{v}_c$. The CoM frame kinetic energies are thus $K_{s/b}^{(1,\text{cm})}=(m_{b/s}/M)K_{tot}^{(1,\text{cm})}$, where
\begin{equation}
    K_{tot}^{(1,\text{cm})}=\frac{1}{M}\left(m_b K_s^{(1,\text{lab})}+m_s K_b^{(1,\text{lab})}-2\mu_{sb}\sqrt{m_s m_b K_s^{(1,\text{lab})}K_b^{(1,\text{lab})}}\right)
\end{equation}
is the total kinetic energy in the CoM frame.

In transforming between frames, we must record the translational kinetic energy of the CoM frame
\begin{equation}
    K_c=\frac{1}{M}\left(m_s K_s^{(1,\text{lab})}+m_b K_b^{(1,\text{lab})}+2\mu_{sb}\sqrt{m_s m_b K_s^{(1,\text{lab})}K_b^{(1,\text{lab})}}\right),
\end{equation}
as well as the direction vector of the transformation. Since we assume an isotropic plasma and do not follow the trajectories of the particles, we only need to know the orientation of the transformation relative to the CoM collision axis. The cosine of the relative angle is $\mu_c=\hat{\mathbf{v}}_c\cdot\hat{\mathbf{u}}_s$, which after some algebra can be expanded to
\begin{equation}
    \mu_c=\frac{\sqrt{m_s m_b}\left(K_s^{(1,\text{lab})}-K_b^{(1,\text{lab})}\right)-\mu_{sb}(m_s-m_b)\sqrt{K_s^{(1,\text{lab})}K_b^{(1,\text{lab})}}}{M\sqrt{K_cK_{tot}^{(1,\text{cm})}}}.
\end{equation}
In the CoM frame, the unit vector for the transformation can be represented with the following convention,
\begin{equation}
    \hat{\mathbf{n}}_c=\left[\mu_c, \sqrt{1-\mu_c^2}, 0\right].
\end{equation}
Finally we can  the CoM-to-lab energy transformation for an outgoing particle,
\begin{equation}
    K_{out}^{(2,\text{lab})}=K_{out}^{(2,\text{cm})}+\frac{m_{out}}{M}K_c+2(\hat{\mathbf{n}}_{out}\cdot \hat{\mathbf{n}}_c)\sqrt{\frac{m_{out}}{M}K_{out}^{(2,\text{cm})}K_c}. 
\end{equation}

\section{Model for p$^{11}$B $\alpha$-particle spectra} \label{appC}

We briefly summarise the physical model from Quebert and Marquez \cite{Quebert69} describing the $\alpha$-particle spectra from the $^{11}$B(p,3$\alpha$) reaction. The physical picture of the three-body breakup is that the excited compound $^{12}$C$^*$ nucleus with energy emits one $\alpha$-particle to leave behind an excited $^8$Be$^*$ nucleus, which then breaks up into two additional $\alpha$-particles. The total kinetic energies before and after the reaction are $K_{tot}^{(1,\text{cm})}$ and $K_{tot}^{(2,\text{cm})}=K_{tot}^{(1,\text{cm})}+Q$, with $Q=8.682$ MeV. The intermediate $\alpha+^8$Be$^*$ state has mean energy $E_0=K_{tot}^{(2,\text{cm})}-Q_2-E^*$, where $Q_2=0.092$ MeV is the reaction energy for the $^8$Be$^*$ breakup and $E^*$ is the $^8$Be$^*$ excitation energy. However, the $^8$Be$^*$ nucleus also has an energy width $\Gamma$ which means the first emitted $\alpha$-particle is not monoenergetic, in contrast to the Monte Carlo model from Hirose \textit{et al.} \cite{Hirose:PPCF25}.

The intrinsic angular momentum states for $^{12}$C$^*$ and $^8$Be$^*$ are $|J^\pi,M\rangle$ and $|L^\pi,\mu\rangle$ respectively, and the orbital angular momentum state of the $\alpha+^8$Be$^*$ system is $|l,m\rangle$. Allowable transitions between states must follow the selection rules,
\begin{gather}
    J=L+\ell, \\
    M=\mu+m,\\
    \pi\left(^{12}\text{C}^*\right)\pi\left(^{8}\text{Be}^*\right)=(-1)^{\ell},
\end{gather}
where the last condition expresses parity conservation.

We denote the kinetic energy and momenta of the $i$th $\alpha$-particle as $\epsilon_i$ and $\mathbf{p}_i$. The spherical coordinate system is defined with polar axis normal to the $\mathbf{p}_1$-$\mathbf{p}_2$ plane and azimuthal reference parallel to $\mathbf{p}_1$. Straightforward kinematics show that
\begin{gather}
    \hat{\mathbf{p}}_i\cdot \hat{\mathbf{p}}_j=\frac{\frac{1}{2}K_{tot}^{(2,\text{cm})} - (\epsilon_i+\epsilon_j)}{\sqrt{\epsilon_i\epsilon_j}} \equiv \cos\theta_{ij}, \\
    \hat{\mathbf{p}}_i\cdot \hat{\mathbf{q}}_{jk}=\frac{K_{tot}^{(2,\text{cm})} - (\epsilon_i+2\epsilon_j)}{\sqrt{\epsilon_i\left(2K_{tot}^{(2,\text{cm})}-3\epsilon_i\right)}} \equiv \cos X_{jk},
\end{gather}
where $\mathbf{q}_{jk}=\mathbf{p}_j-\mathbf{p}_k$ and $K_{tot}^{(2,\text{cm})}$. The azimuthal angles of the three $\alpha$-particles are then $\phi_1=0$, $\phi_2=\theta_{12}$ and $\phi_3=2\pi - \theta_{23}$. For the subsystem in which $\alpha_i$ is emitted first, the partial decay amplitude is given by
\begin{equation}
    m_{i,jk}(M)=\sum_{m=-\ell}^\ell \langle JM|L\mu,\ell m\rangle Y_\ell^m\left(\frac{\pi}{2},\phi_i\right)Y_L^\mu\left(\frac{\pi}{2},\phi_i + X_{jk}\right)F_{L\ell}(\epsilon_i).
\end{equation}
$\langle JM|L\mu,\ell m\rangle $ is the Clebsch-Gordan coefficient and $Y_\ell^m$ and $Y_L^\mu$ are spherical harmonics in the directions of $\mathbf{p}_i$ and $\mathbf{q}_{jk}$, respectively. The factor $F_{L\ell}$ represents the transition from the intermediate $\alpha+^8$Be$^*$ to the final state and is modelled as a Breit-Wigner propagator,
\begin{equation}
    F_{L\ell}(\epsilon_i)=\frac{1}{\frac{3}{2}\epsilon_i-E_0+\frac{1}{2}i\Gamma}.
\end{equation}

Since $m_{i,jk}=m_{i,kj}$, there are only three unique indistinguishable decay sequences which we must sum together to compute the total decay amplitude,
\begin{equation}
    A(M)=m_{1,23}(M) + m_{2,31}(M) + m_{3,12}(M),
\end{equation}
for which the transition matrix element follows from
\begin{equation}
    \frac{d^2\sigma}{d\epsilon_1 d\epsilon_2}\propto |\mathcal{M}_{if}|^2=\sum_{M=-J}^J |A(M)|^2.
\end{equation}

A key limitation of this model is that we can only consider one type of decay channel. We choose this to be the broad resonance at $K_{tot}^{(1,\text{cm})}=620$ keV, which proceeds from a $^{12}$C$^*(2^-)$ state to a $^{8}$Be$^*(2^+)$ with $E^*=3$ MeV and $\Gamma=1.45$ MeV. We take $\ell=3$, in line with both the original paper as well as the latest cross-section model \cite{Sikora:JFE16}.

\bibliographystyle{apsrev4-2}
\bibliography{refs}

\end{document}